\documentclass[prl,twocolumn,showpacs,floatfix,preprintnumbers,amsmath, amssymb,superscriptaddress]{revtex4} 

\usepackage{graphicx}
\usepackage{amsmath}
\usepackage{amssymb}
\usepackage{color}
\usepackage{dcolumn}
\usepackage{epsfig}
\usepackage{bm}
\usepackage[urlcolor=blue]{hyperref}
\hypersetup{backref, colorlinks=true, linkcolor=blue, citecolor=blue}

\begin{document}

\title{Prediction of Superconductivity in Potassium-Doped Benzene}

\author{Guohua Zhong}

\affiliation{Center for Photovoltaics and Solar Energy, Shenzhen Institutes of Advanced Technology, Chinese Academy of Sciences and The Chinese University of Hong Kong, Shenzhen 518055, China}
\affiliation{Beijing Computational Science Research Center, Beijing 100089,
China}

\author{Xiao-Jia Chen}
\email{xjchen@hpstar.ac.cn} \affiliation{Center for High Pressure Science and
Technology Advanced Research, Shanghai 201203, China}

\author{Hai-Qing Lin}
\email{haiqing0@csrc.ac.cn} \affiliation{Beijing Computational Science Research
Center, Beijing 100089, China}
\affiliation{Center for Photovoltaics and Solar Energy, Shenzhen Institutes of
Advanced Technology, Chinese Academy of Sciences and The Chinese University of
Hong Kong, Shenzhen 518055, China}

\date{\today}

\begin{abstract}
To explore underlying mechanism for the superconducting phase in recent discovered aromatic hydrocarbons, we carry out the first-principles calculations on benzene, the basic and the simplest unit of the series and examine the structural and phase stability when doped by potassium, K$_x$C$_6$H$_6$ ($x=1,2,3$). We find that K$_2$C$_6$H$_6$ with the space group of $Pbca$ is the most stable phase with superconducting transition temperature around 6.2 K. Moreover, we argue that all existing hydrocarbons should have a unified superconducting phase in the same temperature range of 5$-$7 K, when doped by two potassium atoms. Our results indicate that the electron-phonon interaction is enough to account for the superconductivity of this unified superconducting phase.
\end{abstract}


\maketitle
Organic superconductors are unique materials with a crystal structure made primarily of a complex carbon based network. Carbon, an element associated directly with life, which was postulated to have a high transition temperature even above room temperature, from a theoretical viewpoint \cite{ref1}. Organic superconductors are generally charge-transfer compounds that are made up of electron donor molecules such as tetrathiafulvalene (TTF), bis-ethylenedithrio-TTF, (BEDT-TTF, abbreviated as ET), and tetramethyltetraselenafulvalene (TMTSF) derivatives, as well as electron acceptor molecules such as tetracyanoquinodimethane (TCNQ) or fullerides \cite{ref2,ref3,ref4,ref5,ref6}. The recent discovery of superconductivity in polycyclic aromatic hydrocarbons (PAHs) \cite{ref7,ref8,ref9,ref10} has led to a great interest in these organic based compounds. Increasing the number of benzene rings, thus delocalizing the charge over a larger space, superconducting transition temperature, $T_c$, could be higher than 30 K and thus revitalizing the hope of realizing high-$T_c$ superconductivity in light element materials. However, there are issues to be addresses such as what the superconducting phase of these hydrocarbons is? how many carries are needed to induce superconductivity? and where the doped carriers are distributed in the unit cells, etc. Having such information is important not only for understanding the physical properties and mechanism of superconductivity in these materials but also for designing materials with higher $T_c$ in technology applications.

The simplest aromatic hydrocarbon is nothing but the benzene, which has six carbons on a ring and six hydrogens attached, C$_6$H$_6$. Other PAHs are made of more rings of carbons and hydrogens. Being the basic cell of PAH, solid benzene stands out as an ideal system to address all concerned issues mentioned above. A fundamental question is whether such a system could superconduct when doped by alkali atoms. If it superconducts, will its properties be of common features for all aromatic hydrocarbons? Thus, explore structures of doped solid benzene and examine corresponding phase stability, followed by the study of superconductivity, is the task of this work.

Solid benzene has rich phase diagram with several phases as functions of temperature and pressure, whose crystallization characteristics has been extensively investigated \cite{ref11,ref12,ref13,ref14,ref15,ref16,ref17}. Liquid benzene crystallizes at room temperature and about 0.07 GPa, in an orthorhombic phase with space group $Pbca$ (phase I), containing four formula units (f.u.) per unit cell \cite{ref11}. At 1.4 GPa, a transition occurs from phase I to phase II ($P$4$_3$2$_1$2), and then at 4 GPa, phase II transfers to phase III ($P$2$_1$/$c$), followed by phase IV (11 GPa) and phase V at high pressures. Solid benzene is non-metallic and non-superconductive in a very large range of pressure. Upon heavy compression, solid benzene was predicted to become a metal from insulator in the pressure range of $180-200$ GPa \cite{ref18}, which means that benzene is a possible superconductor under high pressure. However, extreme pressure condition brings the difficulty to experimental fabrication and observation. Hence, we choose potassium (K) to dope solid benzene at ambient pressure to explore the realization of metallization and superconductivity. If doped solid benzene superconducts at ambient pressure, it would be possible to have higher $T_c$ at high pressure.

Although experimental studies on doping K into solid benzene have not been performed yet, the interaction between benzene molecule and K$^{+}$ ion has been observed experimentally and studied theoretically by E. L\'{o}pez \emph{et al}. \cite{ref19}. Their work implies the feasibility of doping potassium into solid benzene. Notably, doping potassium has been shown to be the most effective and common method to induce superconductivity in PAHs \cite{ref7,ref8,ref9,ref10}: all aromatic hydrocarbons were found to become superconductor with K dopant. One reason is that K has a relatively low melting point as compared to other alkali elements, which enables the chemical reaction and crystallization from liquid states of K and benzene easy to occur.

For doped solid benzene, we firstly perform structure optimization to obtain its atomic positions and lattice constants. Then we calculate corresponding electronic structures and electron-phonon interactions. All structural optimizations and electronic properties were calculated by using the Vienna \emph{ab} initio simulation package (VASP) \cite{ref20}. In this work, the projector augmented wave (PAW) method \cite{ref21} was adopted, and the exchange correlation energy was described by the Ceperley-Alder local density approximation (LDA) \cite{ref22} as parameterized by Perdew and Zunger \cite{ref23}, which has been confirmed to be adequate by previous studies \cite{ref24,ref25,ref26,ref27,ref28}. An energy cutoff of 500 eV was used for the plane wave basis sets. The Brillouin zone (BZ) was sampled by $4\times4\times4$ Monkhorst-Pack \emph{k}-point grids during the optimization, and double \emph{k}-point in the electronic structural calculations. The convergence thresholds were set as 10$^{-6}$ eV in energy and 0.005 eV/{\AA} in force. A conjugate-gradient algorithm was used to relax the ions into their instantaneous ground state.

To calculate electron-phonon interaction and relevant parameters, we use the QUANTUM-ESPRESSO package (QE) \cite{ref29} with a cutoff energy of 60 and 450 Ry for wave functions and charge densities, respectively. Forces and stresses for the converged structures were optimized and checked to be within the error allowance of the VASP and QE codes. Troullier-Martins norm-conserving scheme \cite{ref30} was used to generate the pseudopotentials for C, H, and K atoms. $8\times8\times8$ Monkhorst-Pack \emph{k}-point grid with Gaussian smearing of 0.03 Ry was used for the electron-phonon interaction matrix element calculations at $2\times2\times2$ \emph{q}-point mesh.

\begin{figure}
\includegraphics[width=\columnwidth]{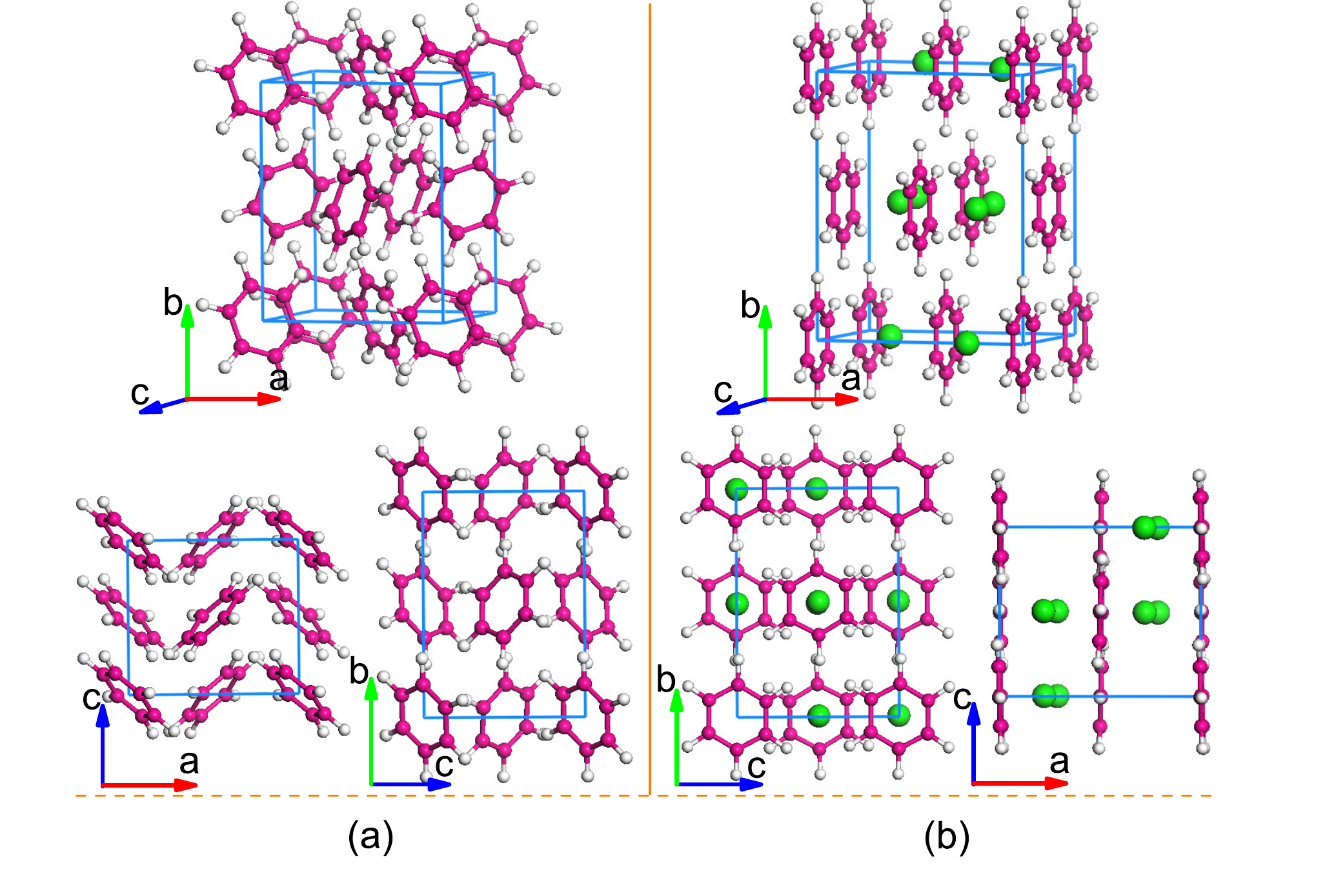}
\caption{(Color online) Optimized crystal structures of (a) solid benzene and (b) K-doped solid benzene, K$_2$C$_6$H$_6$, viewing from different directions. Purple balls represent K atoms.}
\end{figure}

As a comparison and a check on accuracy, we start our investigation by looking at the solid benzene. Figure 1(a) shows the optimized crystal structure of solid benzene viewing from different directions, with crystal lattice constants being $a=7.041$ {\AA}, $b=8.903$ {\AA}, and $c=6.357$ {\AA} at zero temperature and the ambient pressure, which are 3-6\% less than experimental values at 78 K \cite{ref11}. Considering temperature effect, this error between theoretical prediction and experimental measurement is acceptable. From reported data on PAHs, the superconductivity is highly sensitive to both the doping concentration and the cation positions in the unit cell. To examine both effects, we perform optimization studies on doped solid benzene, K$_x$C$_6$H$_6$, by varying potassium contents ($x$ = 1, 2, 3) and position of K atoms in the unit cell. To check their stability, formation energy, as defined by $E_{formation} = E_{K_xC_6H_6} - E_{C_6H_6} - E_{K_x}$, is a good indication. We obtained the results of formation energy as 0.35, $-2.99$, and 1.52 eV for $x$ = 1, 2, and 3, respectively. Thus, both K$_1$C$_6$H$_6$ and K$_3$C$_6$H$_6$ are unstable and the K$_2$C$_6$H$_6$ is the only stable phase in this series. Its big negative formation suggests easier synthesization.

Figure 1(b) shows the optimized K$_2$C$_6$H$_6$ crystal structure viewing from different directions, with lattice constants being $a=8.605$ {\AA}, $b=10.705$ {\AA}, and $c=6.415$ {\AA}, respectively. Comparing with the undoped case, the lattice constants expand largely in the \emph{a} and \emph{b} directions. The inserting of eight K atoms makes the volume increased by 26.9\%.  Under K doping, a visible layer shape structure is formed, differing from that of the solid benzene. As shown in Fig. 1(b), all benzene rings are rotated and parallel to the \emph{bc}-plane under the interaction of dopants, and K atoms are intercalated into two layers of benzene rings, occupying on Wyckoff $8c$ sites. We found that the optimized structure of K$_2$C$_6$H$_6$ has tended to the higher symmetry of $Fmmm$ from the lower $Pbca$ symmetry. Significantly, the layered structure of K$_2$C$_6$H$_6$ is similar to those of YbC$_6$ and CaC$_6$ \cite{ref31}. This similarity indicates the possibility of superconducting, because it is known that both YbC$_6$ and CaC$_6$ are superconductors, with the $T_c$ reaching to 6.5 K and 11.5 K, respectively \cite{ref31,ref32}.

\begin{figure}
\includegraphics[width=\columnwidth]{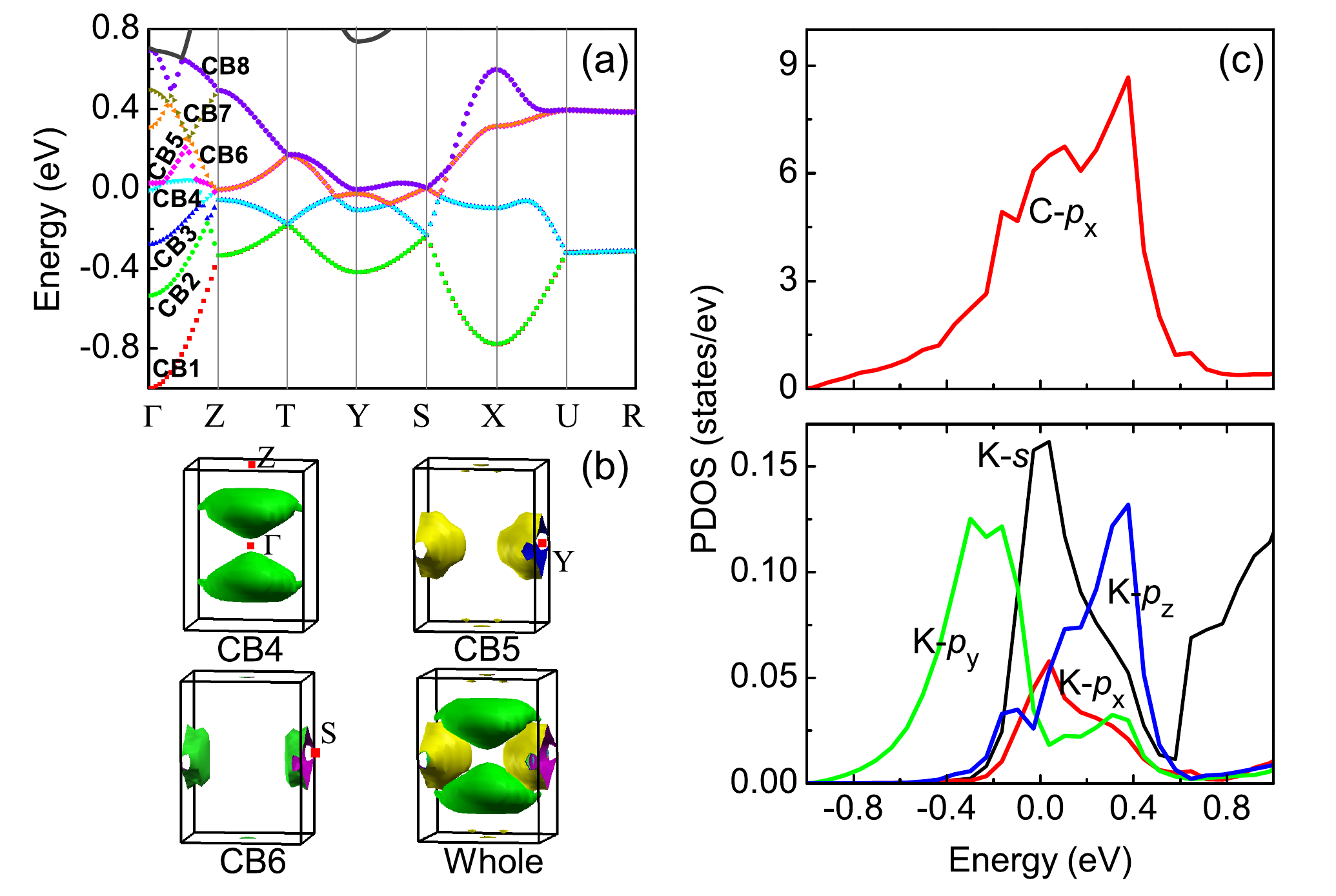}
\caption{Electronic band structure of K$_2$C$_6$H$_6$: (a) energy band; (b) Fermi surface; and (c) projected density of states (PDOS), where zero energy denotes the Fermi level.}
\end{figure}

Once the crystal structure is determined, it is ready to obtain the electronic band structure. For K$_2$C$_6$H$_6$, we show 8 conduction bands (called as CB1, CB2, ..., and CB8) along a high symmetry $k$-direction, $\Gamma-Z-T-Y-S-X-U-R$, of the Brillium zone in Fig. 2(a). Specifically, under either \emph{Pbca} or \emph{Fmmm} symmetry, the LUMO and LUMO+1 orbitals split into 8 bands from $\Gamma$ to $Z$, then they merge into 4 degenerate bands along the $Z-T-Y-S-X-U$ directions, and further into two highly degenerated bands along the $U-R$ direction. Electrons transferred from the K atoms to the C$_6$H$_6$ molecule, occupying low energy orbitals and raise the Fermi level. Three bands, CB1, CB2, and CB3 are fully occupied by six of eight electrons, while the other three, CB4, CB5, and CB6 are partially filled by the remaining two electrons with Fermi level passing through. Both CB7 and CB8 are empty. Thus, K$_2$C$_6$H$_6$ is metallic and it is in the low-spin state. Comparing with the solid benzene, the intermolecular distance along benzene rings becomes smaller on the \emph{bc}-plane of K$_2$C$_6$H$_6$, which enhances the intermolecular interaction and broadens the dispersion of energy bands. Analyzing the electronic characters near the Fermi level, as shown by Fig. 2(b), the hole-like Fermi surfaces are formed by CB4 along the $\Gamma-Z$ \emph{k}-direction, while both CB5 and CB6 contribute the electron-like Fermi surfaces around the $Y$ \emph{k}-point. Fermi surface nesting is quite clear. Calculation of the projected density of states (PDOS) show that C-$p_y$ and C-$p_z$ states are fully occupied and far below the Fermi level, while C-$p_x$ state has a big PDOS around the Fermi level, as shown in Fig. 2(c). Most PDOS of K-($s, p_x, p_y, p_z$) states are above the Fermi level, which indicates that the electrons transfer from K to C-$p_x$ orbital. The large ratio of C-$p_x$ PDOS over K PDOS shows that C-$p_x$ electronic states contribute more to the Fermi surfaces.

\begin{figure}
\includegraphics[width=\columnwidth]{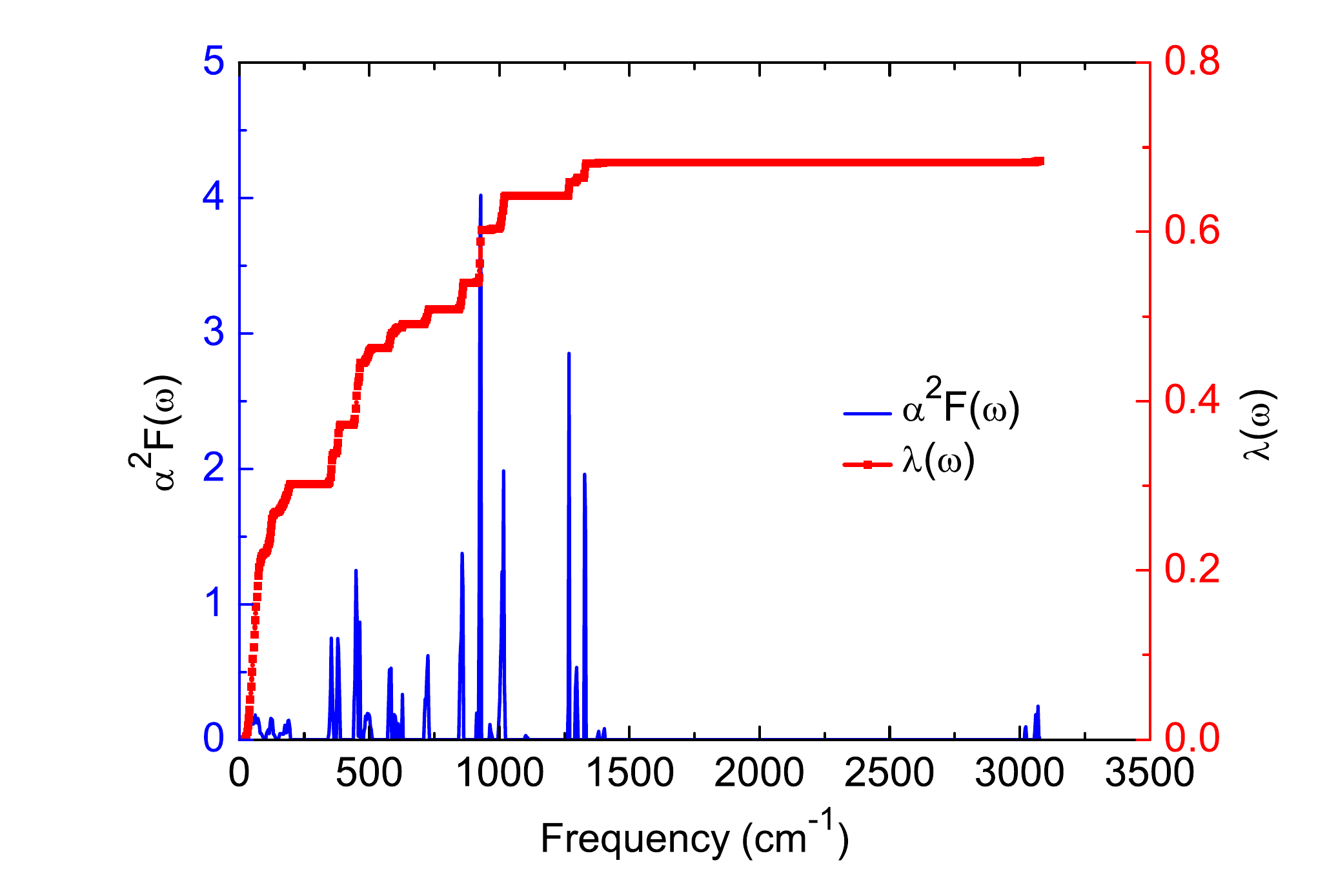}
\caption{Electron-phonon coupling of K$_2$C$_6$H$_6$: Eliashberg spectral function $\alpha^2F(\omega)$ and the electron-phonon coupling integral
$\lambda(\omega)$.}
\end{figure}

Now we examine whether this simplest aromatic hydrocarbon could exhibit superconductivity like other PAHs wiith long benzene rings \cite{ref7,ref8,ref9,ref10}. Assuming the pairing mechanism is still the electron-phonon interaction mediated, we evaluate the Eliashberg spectral function $\alpha^{2}F(\omega)$ and the electron-phonon coupling integral $\lambda(\omega)$ of K$_2$C$_6$H$_6$ as a function of frequency. The results are shown in Fig. 3. The vibration in the low frequency range of $0-250$ cm$^{-1}$ comes from K atoms and intermolecular modes, while the vibration with moderate frequencies of $300-1450$ cm$^{-1}$ mainly comes from C atoms. The total $\lambda$ is 0.67, which is comparable to that of K$_3$C$_{22}$H$_{14}$ \cite{ref25}. We also noted that vibrational modes from C atoms contribute 56\% to the total $\lambda$ value, and the rest 44\% is provided by the dopant and intermolecular phonon modes. Following the modified McMillan equation of Allen and Dynes \cite{ref34}, with the obtained values of $\lambda(\omega)$ and $\omega_{log}$ for K$_2$C$_6$H$_6$, and a typical value of 0.10 for the Coulomb pseudopotential $\mu^{\star}$, we obtain $T_c$ = 6.2 K. The choice of $\mu^{\star}$ is appropriate for light-element systems \cite{ref25,ref35,ref36}.

We want to point out that the dominant contribution to the total $\lambda$ from the vibrational modes of the C atoms in K$_2$C$_6$H$_6$ shows some similarity with other C$_6$-based or low carbon-bearing compounds. In reality, CaC$_6$ was discovered to exhibit superconductivity at 11.5 K \cite{ref31}. Superconductivity has also been discovered in intercalation compounds of graphite with alkali metals $A$C$_8$ ($A =$K, Rb, or Cs) \cite{ref37}. The predicted $T_c$ of 6.2 K in K$_2$C$_6$H$_6$ is comparable to the temperature range of these low carbon-based superconductors \cite{ref31,ref37}.

A comparison of the predicted superconductivity in benzene with other aromatic hydrocarbons reveals a common feature of PAH (Fig. 4). Besides the reported high-$T_c$ phases with $T_c=18$ K for K-doped picene (K$_x$C$_{22}$H$_{14}$) \cite{ref7}, 15 K for K-doped coronene (K$_x$C$_{24}$H$_{12}$) \cite{ref8}, and 33 K for K-doped 1,2:8,9-dibenzpentacene (K$_x$C$_{30}$H$_{18}$) \cite{ref10}, multi-superconducting phases in long-benzene-ring hydrocarbons were generally observed. There exists a low $T_c$ phase with almost the same $T_c$ between $5-7$ K in these hydrocarbon superconductors discovered so far \cite{ref7,ref8,ref9,ref10}: K$_x$C$_{14}$H$_{10}$ has $T_c$ 5 K \cite{ref9}; K$_x$C$_{22}$H$_{14}$ has $T_c$ 7 K \cite{ref7,ref8}; K$_x$C$_{30}$H$_{18}$ has $T_c$ of 7.4 K \cite{ref10}, though several other transitions with higher $T_c$'s were observed. Moreover, Rb$_3$C$_{22}$H$_{14}$ and Ca$_{1.5}$C$_{22}$H$_{14}$ both show transition temperature around 7 K \cite{ref7}. $T_c$ of $5-7$ K was reported for K$_x$C$_{24}$H$_{12}$ \cite{ref8}, in addition to other three more transitions. Superconductivity at 6.4 K was observed in Sm$_{1.25}$chrysene \cite{ref38}. It is important to note that only in the range of $5-7$ K transition temperatures were observed in all reported superconducting aromatic hydrocarbons with various number of benzene rings, in a good agreement with our current prediction for benzene, the simplest compound of this series. In fact, the value of the logarithmic average of phonon frequencies $\omega_{log}$ (199.8 K) for K$_2$C$_6$H$_6$ is exactly the same as that of K$_3$C$_{22}$H$_{14}$ \cite{ref25}, seemingly independent of the number of the benzene rings. A careful analysis on the 2-ring case also showed that a possible superconducting phase with $T_c$ in the same range for K$_2$C$_{10}$H$_8$. With these analysis, we are confident to conclude that the presence of the superconducting phase with $T_c$ around $5-7$ K is a common feature of aromatic hydrocarbons. In most cases, the low-$T_c$ phase was easily fabricated and more stable than the higher $T_c$ phase \cite{ref7,ref8,ref10}. Considering the fact that the cation concentration has never been experimentally identified by using the reliable neutron scattering or single crystal X-ray diffraction techniques, all the reported doping levels are an ideal chemical estimation based on the synthesis procedure. Our phase stability examination for potassium-doped benzene pins down the doping concentration of the exact two electrons for the stabile superconducting phase of these hydrocarbons with $T_c$ of $5-7$ K.

\begin{figure}
\includegraphics[width=\columnwidth]{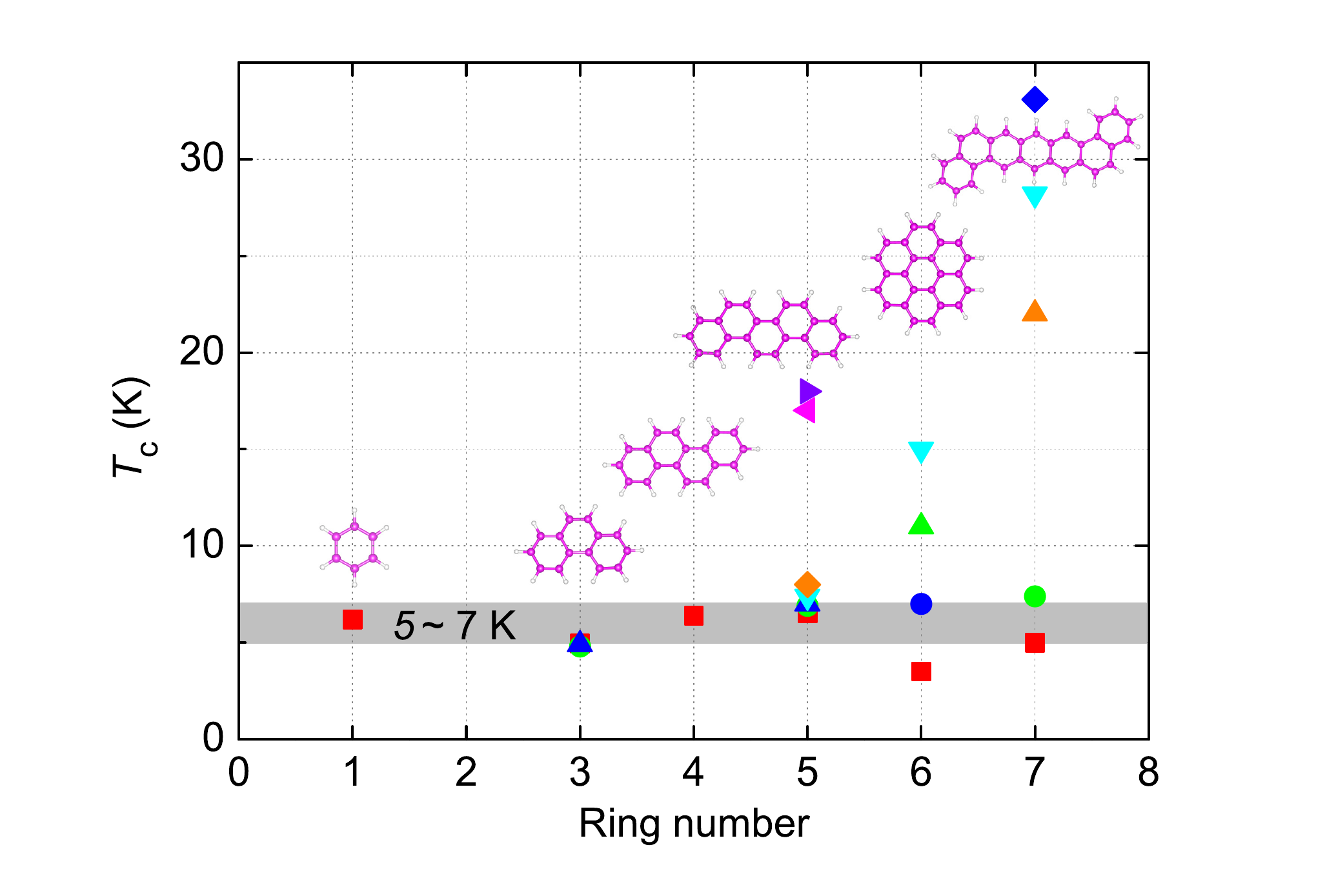}
\caption{Variation of superconducting transition temperature $T_{c}$ with the number of benzene rings for various aromatic hydrocarbons. The solid red square is the prediction of this work for benzene. Other symbols are taken from experiments \cite{ref7,ref8,ref9,ref10,ref38}. The dashed line is the guide for eyes.}
\end{figure}

Although a complete understanding of the mechanism of superconductivity in aromatic hydrocarbons has yet to achieve, the common feature of the $5-7$ K superconducting phase for all reported aromatic hydrocarbons may provide us some clues. Moreover, further investigations indicate that the electron-electron correlations seem to play an important role on superconductivity in these systems. This is quite similar to charge-transfer organic superconductors and fullerides \cite{ref5,ref40,ref41,ref42,ref43,ref44,ref45} where the importance of electron correlations on superconductivity was emphasized. Magnetic measurements \cite{ref9} on phenanthrene suggest the existence of local spin moments, and indicate similar interplay between magnetism and superconductivity observed in fullerides \cite{ref5,ref40,ref41,ref42,ref43,ref44,ref45}. We indeed found that the electron-electron correlation becomes important as the number of benzene rings increases. Multi-ring hydrocarbons are in favor of strong correlations and thus tend to have higher $T_c$'s. It is therefore believed that both the electron-phonon interaction and electron-electron correlations work together to enhance $T_c$ of the aromatic hydrocarbons with increasing the number of benzene rings.

In summary, carrying out the first-principles calculations, we have investigated the crystal structures and possible superconductivity in potassium doped solid benzene, K$_x$C$_6$H$_6$ ($x=1,2,3$). The K dopant results in the visible change of crystal configuration and drives the system from insulator to metal. K$_2$C$_6$H$_6$ is found to be stable due to its lowest formation energy as compared to other doping. Superconductivity in doped solid benzene with a $T_c$ of 6.2 K is predicted. This superconducting phase shares the common feature of all existing aromatic hydrocarbon superconductors.

The work was supported by the Natural Science Foundation of China (Grant Nos. 11274335, 91230203, and U1230202), and the Shenzhen Basic Research Grant (Nos. KQC201109050091A and JCYJ20120617151835515).

\bibliography{aipsamp}

\end{document}